\documentclass[10pt,journal,compsoc]{IEEEtran}

	\usepackage{booktabs}

	\newcommand{\reffig}[1]{Fig.~\ref{#1}}
	
	\newcommand{\reftbl}[1]{Table~\ref{#1}}
	\newcommand{\refsec}[1]{Sec.~\ref{#1}}

	\newcommand{\hAbs}[1]{\ensuremath{\left \lvert \, #1 \, \right \rvert} } 

	\usepackage[utf8]{inputenc} 
	\usepackage[iso]{datetime}
	\usepackage{amsmath,amssymb,amsfonts,amsthm}
	\usepackage{enumerate}
	\usepackage{multicol}

	\usepackage{xcolor}
		\definecolor{darkred}{rgb}{0.8,0.1,0.1}
		\definecolor{darkgreen}{rgb}{0,0.5,0}
		\definecolor{darkblue}{rgb}{0,0,0.5}
		\colorlet{RED}{red}
	
	\usepackage[colorlinks=true,linkcolor=red,urlcolor=blue,citecolor=red]%
		{hyperref}
	\usepackage{graphicx,epstopdf}
	\graphicspath{{../common/figures/}}
	\usepackage{subcaption}

\usepackage{url}

\begin{document}

\title{
	Optimism brings accurate perception\\ 
	in iterated prisoner's dilemma
}

\author{
	Orhun~Görkem
	and
	Haluk~O.~Bingol
	\IEEEcompsocitemizethanks{
		\IEEEcompsocthanksitem O.~Görkem and H.~O.~Bingol  are with Bogazici University.
	}%
}

\IEEEtitleabstractindextext{%
	\begin{abstract}
		We analyze an extended model of the Iterated Prisoner's Dilemma 
		where agents decide to play 
		based on the data from their limited memory or recommendations. 
		The cooperators can decide whether to play 
		with the matched opponent or not. 
		The agents' decisions are directly linked to their level of optimism since they decide to play 
		if they believe the opponent has a high probability of cooperating.
		Optimism is precisely tuned by parameters optimism threshold and tolerance. 
		Our experiment showed that being optimistic is better for cooperators
		as it leads to more accurate exploration in the multi-agent system, 
		which tolerates the vulnerability against defectors.
	\end{abstract}
	\begin{IEEEkeywords}
		Complex networks; iterated prisoner's dilemma; tolerance; optimism; Multi-agent systems; misjudgment; exploration
	\end{IEEEkeywords}
}

\maketitle

\section{Introduction}

With the exponential population growth, 
large societies with frequent interactions between strangers emerged~\cite{%
	durand1977}. 
This brought a necessity to analyze complex multi-agent systems 
to recognize the patterns of social and economic behaviors. 
Many directed network models are developed to investigate 
the social influence and trust propagation among agents.
Ref~\cite{%
	guha2004propagation} 
reports that trust between any pair of agents in a huge population can be 
accurately predicted with even a small number of trusts given as input. 
This recursive effect of trust has been used in social recommendation networks such as DiffNet~\cite{%
	wu2019}.  
Using the trusted links is called exploitation. 
However, agents sometimes benefit from exploration instead of exploitation by taking the risk of forming new trust links with new agents.

This brings the debate to the optimality in the tradeoff between exploration and exploitation. 
Multi-armed bandit problem is studied, and Upper Confidence Bound algorithm is
designed to achieve analytically solid performance bounds against this tradeoff~\cite{%
	auer2000}.  
Upper Confidence Bound algorithm invokes a formula to pick the next action to be taken. The formula has a term for uncertainty, which increases when the action is not being selected for a while. 
This makes the unselected action more likely to be selected. 
The incentive for exploration is analytically shown by the Upper Confidence Bound algorithm. 
Statistical evidence in favor of exploration is also reported in a randomized field experiment 
	where participants are suggested to make life decisions according to coin tosses \cite{%
	levitt2016}. 
The study reveals that individuals who are told by the coin toss to make a change in their life are happier six months later 
	than those who were told by the coin to maintain the status quo \cite{%
	levitt2016}.

On the other hand, Prisoner's Dilemma(PD) is one of the most popular models 
used to examine the dynamics of multi-agent systems. 
For instance, PD models are used to test the effect of economic incentives to increase cooperation~\cite{%
	schweitzer2020}, 
and to test the effects of recommendation in social networks~\cite{%
	cinar2020getting}. 
By changing the capabilities of agents, and the type of information that agents receive, 
different settings can be observed in Prisoner's Dilemma models. 
Ref~\cite{%
	schweitzer2013}
reveals that agents tend to choose cooperation collectively 
if the payoff information is not shared with them. 
Ref~\cite{%
	cetin2014}
introduces limited attention concept with a choice-refusal opportunity to cooperators 
in Iterated Prisoner's Dilemma to infer that cooperators benefit from the increase in memory 
more than defectors. 
Finding in~\cite{%
	cetin2014}
is extended with the results from~\cite{%
	cetin2016dose}
as payoff matrix is found to be effective in the evolution of memory. 
In another setting,~\cite{%
	cinar2020getting}
shows that getting recommendations from other agents may not be useful 
for cooperators in case of insufficiency in memory. 

Our work models a specific version of Iterated Prisoner's Dilemma 
to view a direction to optimal tradeoff between exploration and exploitation in multi-agent systems. 
We parametrize optimism to yield better exploration 
for cooperators to detect defectors more accurately. 
We claim that: 
\begin{itemize}
     \item Cooperators should keep playing with other agents to some extent, 
     even if those agents are considered to be defectors.
     \item Playing one game with an agent is insufficient for accurate exploration. 
\end{itemize}

\section{Background}

\subsection{Prisoner's Dilemma}

Prisoner's Dilemma (PD) is a simple game to evaluate the success of various game strategies by trying them against an opponent whose acts are unknown and unpredictable~\cite{%
	axelrod1981evolution}. 
Two agents playing Prisoner's dilemma are expected to choose between defecting and cooperating. 
According to their choices, they are awarded or punished.
If an agent cooperates while the other defects, the cooperator gets the \emph{sucker} payoff $S$ and 
the defector gets the \emph{temptation} payoff $T$. 
If both players choose to cooperate, 
then they both get the \emph{reward} payoff $R$. 
In the case of mutual defection, both players get the \emph{punishment} payoff $P$. 
In Prisoner’s Dilemma game, the payoffs should satisfy
both $S < P < R < T$ and $S + T < 2R$~\cite{%
	axelrod1981evolution}.

For one round, it is always better to defect without regarding the act of the opponent agent 
if there is no specific modification as in The Expected Prisoner's Dilemma~\cite{%
	arend2020}.
However, in real life, we come up against the consequences of our previous acts. 
Therefore, \emph{Iterated Prisoner's Dilemma} (IPD) is introduced. 
Iterated Prisoner's Dilemma models are often used to explain sentient human behaviors~\cite{%
	press2012iterated}. 
Agents play with each other consecutively, 
and the result is determined according to the sum of consecutive rounds. 
Agents can remember their opponents' previous decisions. 
Then they decide by taking the past rounds into account. 
This generally promotes the power of cooperation since $P < R$ and 
reciprocal trust yield better results than reciprocal distrust. 
Yet, IPD with $N$ people where $N>2$ is better than IPD with 2 people 
in the aspect of representing some of the real-world problems 
such as \emph{conversation of scarce resources} 
and 
\emph{tragedy of commons}~\cite{%
	colman1982,%
	hardin1968}. 
Hence we use a system with $N$ agents in our experiments.

\subsection{IPDwRec}  

Iterated Prisoner’s Dilemma with Recommendation (IPDwRec)~\cite{%
	cinar2020getting}
is a model developed to simulate an environment,
where agents play with each other in several combinations and make use of recommendations in their choices. 
We built our model over IPDwRec by appending some additional terms and procedures.

\subsubsection{Population}

Iterated Prisoner's Dilemma has some well-known strategies such as TitForTat, Grim, and Pavlov~\cite{%
	hingston2009iterated}. 
In these strategies, agents have total control of their behaviors. 
However, IPDwRec is different. 
The population of IPDwRec consists of $N$ self-interested agents~\cite{%
	shoham2008}. 
Half of the agents are cooperators, and the other half are defectors. 
Agents are not pure cooperators or defectors. 
They follow a mixed strategy.  
With the probability of $\epsilon < 0.5$, they act opposite to their characteristic. 
Hence, 
a \emph{cooperator} is expected to defect with probability of $\epsilon$.
Similarly,   
a \emph{defector} is expected to cooperate with probability of $\epsilon$.

\subsubsection{Memory, Perception, and Forgetting}  

Agent $i$ keeps its previous encounters with agent $j$ in its memory.
This data consists of $(c_{ij},d_{ij})$ pair,
where
$c_{ij}$ is the number of times that $j$ cooperated against $i$ and 
$d_{ij}$ is the number of times that $j$ defected against $i$. 
Derived from Laplace's Rule of Succession~\cite{%
	jaynes2003probability},
agent $i$ calculates \emph{perceived cooperation probability}, 
defined as
\begin{align}
	t_{ij} = \frac
		{c_{ij} + 1}
		{c_{ij} + d_{ij} + 2}.
	\label{eq:t_ij}
\end{align}
Perceived cooperation probability is a measure of one-sided trust between agents.
An agent $i$ is said to perceive an agent $j$ as a \emph{cooperator} if $t_{ij} > 0.5$.
Otherwise, $j$ is perceived as a \emph{defector}.   

Agents have a fixed memory size of $M \le N$. 
This means that an agent can store data about $M$ agents in the population. 
\emph{Memory ratio} is defined as $\mu = M / N$.
Suppose $i$ plays with $j$.
If $j$ is already in the memory, $i$ simply updates $c_{ij}$ and $d_{ij}$ values.
If $j$ is not in the memory, a new $(c_{ij}, d_{ij})$ pair is inserted.

Assuming that memory size is not large enough to keep all opponents, i.e., $\mu < 1$,
eventually, the memory of agent $i$ will be full.
In order to keep $j$ in memory, $i$ needs to ``forget'' some other agent.
Therefore a forgetting mechanism is required.
There are three strategies for forgetting: 
(i)~Forget cooperators (FC), 
where the agent to forget is chosen randomly among agents perceived as cooperators, 
(ii)~Forget defectors (FD),
where the agent to forget is chosen randomly among agents perceived as defectors,
(iii)~Forget random (FR),
where the agent to forget is chosen randomly among all agents.

\subsubsection{Decision to Play}

Agents are matched uniformly at random to play with each other. 
A matched agent may refuse to play. 
Cooperators, having a high probability of cooperating, can only protect themselves from defectors 
by refusing to play with them. 
Therefore, they need to be able to detect their opponents' characteristics by their memory or recommendation. 
The procedure of decision to play is the following:  
\begin{enumerate}[i.]

	\item 
	If agent $i$ remembers agent $j$,  it calculates $t_{ij}$. 
	If $j$ is perceived as a cooperator, i.e.,  $t_{ij}>0.5$, 
	$i$ decides to play; otherwise rejects.

	\item 
	If agent $i$ does not remember agent $j$, it inquires recommendation. 
	Agent $i$ calculates $t_{ij}$ with the recommendation procedure explained in \refsec{sec:Recom}. 
	If $t_{ij}>0.5$, 
	$i$ decides to play; otherwise rejects.	
	\item 
	If no recommendation is received, agent $i$ decides to play.
\end{enumerate}
The same procedure applies for $j$ against $i$ too. 
If both agents decide to play, they play the game. 
Otherwise, the game is canceled.

\subsubsection{Recommendation}
\label{sec:Recom}

Recommendation is the most significant feature of IPDwRec when compared to classical IPD. Recommendations are provided with the following procedure:  
\begin{enumerate}[i.]

	\item 
	Agent $i$ is matched with an opponent $j$ who does not exist in its memory.

	\item 
	Then agent $i$ inquires other agents in its memory about agent $j$. 
	The ones 
	who keep information about agent $j$ in their memories 
	are called members of the \emph{set of recommenders} ($R_{ij}$). 
	All agent $k$'s, 
	where $ k \in R_{ij}$, 
	share their perceived cooperation probability ($t_{kj}$) with agent $i$. 
	For instance, if agent $i$ remembers agent $k$ and, agent $k$ remembers agent $j$, 
	then agent $k$ passes $t_{kj}$ to agent $i$. 

	\item 
	Agent $i$ collects all ratios and evaluates them to make a decision.
	The decision depends on dispositions. Agents have three types of innate disposition: 
	(i)~\emph{Optimism},
	where agents take the best recommendation into account, 
	(ii)~\emph{Realism},
	where agents take the mean of recommendations, 
	(iii)~\emph{Pessimism},
	where agents consider the worst recommendation as the cooperation probability. 
		
\end{enumerate}

\section{Proposed Model}

Our model introduces some new features and modifies some rules in IPDwRec~\cite{%
	cinar2020getting} 
to analyze the effect of optimism on exploration more precisely. Note that some of those new features are in favor of defectors.

\subsection{Defectors}

We modify the behavior of defectors.
(i)~In our model, 
defectors always play. 
In IPDwRec~\cite{%
	cinar2020getting}, 
if an agent perceives its opponent as a defector, it rejects to play.
This 
behavior is correct for cooperators but meaningless for defectors. 
For a non-negative payoff matrix,
a defector has nothing to lose 
even if the opponent is a defector~\cite{%
	cetin2016dose}.
(ii)~Moreover, defectors are memoryless as they do not need memory in 
the experimental setup where they always play. 
Note that without memory, defectors do not give recommendations since they keep nothing to share.

With these new rules, 
defectors are more advantageous compared to those in IPDwRec.

\subsection{Optimism Threshold}

In IPDwRec model, agents decide to play 
if the opponent is perceived as cooperator, i.e. 
$t_{ij} > 0.5$.
In our model, 
we relax this by introducing \emph{optimism threshold} $\alpha$.
Then an agent $j$ is perceived as \emph{cooperator} by agent $i$ 
if  $t_{ij} > \alpha_{i}$, and as \emph{defector} otherwise.

By defining optimism threshold, we aimed to adjust the method of asserting dispositions. 
The new model does not include the dispositions of optimist and pessimist of IPDwRec. 
Optimism threshold $\alpha$ determines the disposition. 
For instance, agent $i$ with $\alpha_{i} = 0.3$ is considered to be more optimistic than 
agent $j$ with $\alpha_{j} = 0.5$. 
Note that $\alpha = 0$ for defectors 
since they always play.

\subsection{Recommendation Evaluation}

Since there are no dispositions as optimism, realism and pessimism in this model, 
all agents calculate $t_{ij}$ from recommendation as the following:
\begin{align}
	t_{ij} 
	= 
	\frac
		{\left[ \sum_{k \in R_{ij}} c_{kj} \right] +1}
		{\left[ \sum_{k \in R_{ij}} (c_{kj} + d_{kj}) \right] + 2}.
\end{align}

Note that the algorithm in \refsec{sec:Recom} is modified to apply this formula. Recommender $k$ shares $(c_{kj} , d_{kj})$ pair with agent $i$ instead of $t_{kj}$. 
With this formulation, 
recommendation of the agents 
who played more number of games with agent $j$ 
will have a higher weight.

\subsection{Tolerance}
\label{sec:tol}

Consider the following case.
Agent $i$ does not know $j$.
It inquires for recommendations but receives none.
Hence $i$ decides to play with $j$.
The game takes place and $j$ defects.
Since $j$ is played once,
$j$ is now in the memory of $i$.
Therefore, no more recommendation takes place the next time they are matched. 
Next time when $j$ is matched, 
$i$ decides based on its memory.
In IPDwRec model, as a result of this one game, 
$i$ evaluates $j$ as a defector and never plays with it as long as $j$ stays in its memory.

In our model, we postpone this decision.
We introduce \emph{tolerance} $\beta$ for cooperators,
which is the number of games that cooperators must play with an opponent 
before classifying it as a defector and rejecting to play. 
Until then, the opponent is perceived as \emph{undefined}. 
Note that this does not apply to defectors since they always play. 
Note also that defectors benefit from this.
For instance, 
a cooperator is guaranteed to play with a defector three times if $\beta = 3$.  

Suppose agents $i$ and $j$ have played a game $k$ times, i.e. $c_{ij}+d_{ij} = k$. 
Agent $i$ perceives $j$ as a
\emph{cooperator} if $t_{ij} > \alpha_i \text{ and } k \geq \beta$,
\emph{defector}   if $t_{ij} \leq \alpha_i \text{ and } k \geq \beta$, and
\emph{undefined} if $k < \beta$.

If an opponent is perceived as a cooperator or undefined, it is not rejected. 
Undefined implies that the agent is not ``well-known'' enough, 
therefore, no bias should be shown against it.   

Note that tolerance is not applied 
if at least one recommendation is received,
and  $t_{ij} < \alpha_{i}$. 
This is because recommendation evaluation accumulates information from several agents. 
In this case, the opponent can be considered well-known.

\subsection{Metrics}

\subsubsection{Payoff Ratio}

We use the same metric defined in Ref~\cite{cinar2020getting} for payoff ratio.
Payoff ratio is used to evaluate the performances of cooperators and defectors. 
As stated in IPDwRec model, 
\emph{average payoff} of a set of agents $A$ is defined as 
\begin{align}
	\overline{ P_A} 
	= 
	\frac
		{1}
		{\hAbs{A}}
	\sum_{i \in A} \text{payoff}(i).
\end{align}
where payoff($i$) is the total payoff collected by agent $i$ among all games played by $i$. 
Now call the set of cooperators in the population as $C$ and defectors as $D$. 
We define \emph{payoff ratio} as  
\begin{align}
	\phi 
	= 
	\frac
		{\overline{P_C}}
		{\overline{P_{C \cup D}}},
\end{align}
where $\phi > 1$ means cooperators are more successful than defectors on average.

\subsubsection{Misjudgments}
\label{sec:misjudge}

The metric for misjudgments is similar to the one defined in Ref~\cite{cinar2020getting}.
There are two possible misjudgments made by cooperators: 
(i)~\emph{$\eta_{cd}$}: number of times when cooperator opponent is perceived as a defector, 
(ii)~\emph{$\eta_{dc}$}: number of times when defector opponent is perceived as a cooperator. 
Note that the defectors do not need to judge their opponents since they always play.

\section{Realizations}

We used payoff matrix of $(S, P, R, T) = (0, 1, 3, 5)$ in our simulations.

In each simulation realization, we used population size of $N = 100$,
where 50 agents are cooperators and 50 agents are defectors. 
We set $\epsilon = 0.1$. 
Thus, cooperation probabilities are 0.9 and 0.1 for cooperators and defectors, 
respectively.  

A realization includes $\tau {N \choose 2}$ pairing of agents, 
where two agents are picked uniformly at random and offered to play. 
Therefore, any agent pair is matched to play $\tau$ times on average.
Note that matched agents play if none of the parties reject. 
We choose $\tau=30$ for each realization as in Ref~\cite{%
	cinar2020getting}. 

We call memory ratio $\mu$, optimism threshold $\alpha$, tolerance $\beta$, and forgetting strategy as \emph{model parameters}. 
All cooperators have the same model parameters in a simulation. 
We report the average of $\mathcal{R} = 30$ realizations to narrow down the confidence interval of model responses.

\section{Results and Discussions}

By introducing optimism threshold $\alpha$ and tolerance $\beta$, 
we had the chance to examine new dispositions and their effects on cooperator performances.  
As an overview,
we have the following observations.

\begin{figure*}[tbp!]
\begin{multicols}{2}
	\center
	\vspace{-1.5\baselineskip}
	\begin{subfigure}{.9\columnwidth}%
		\includegraphics[width=\columnwidth]
			{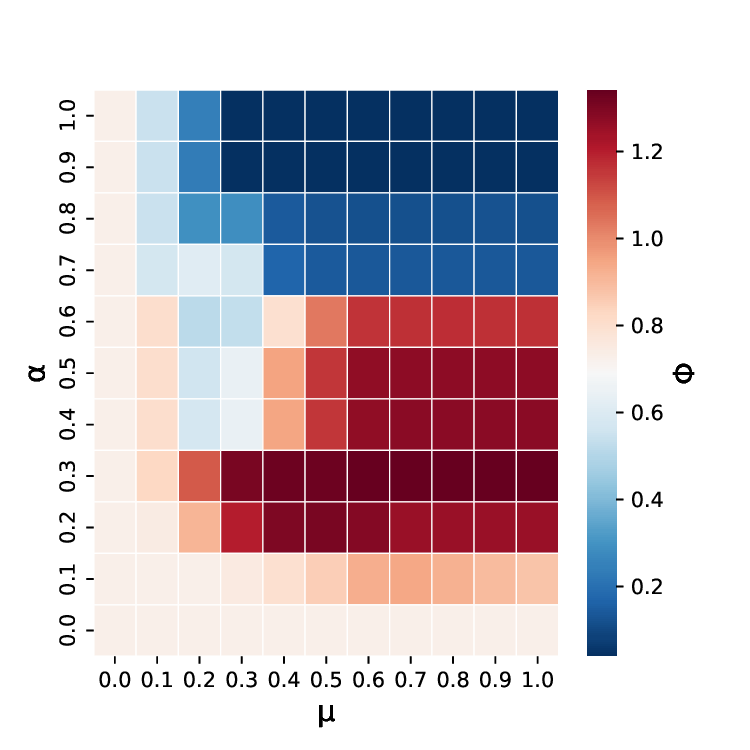}
			\vspace{-1.5\baselineskip}
		\caption{FC, $\beta=1$} 
		\label{fig:FC_PAYOFF}
	\end{subfigure}
	\vspace{-1.5\baselineskip}
	\begin{subfigure}{.9\columnwidth}%
	\vspace{-1.5\baselineskip}
		\includegraphics[width=\columnwidth]
			{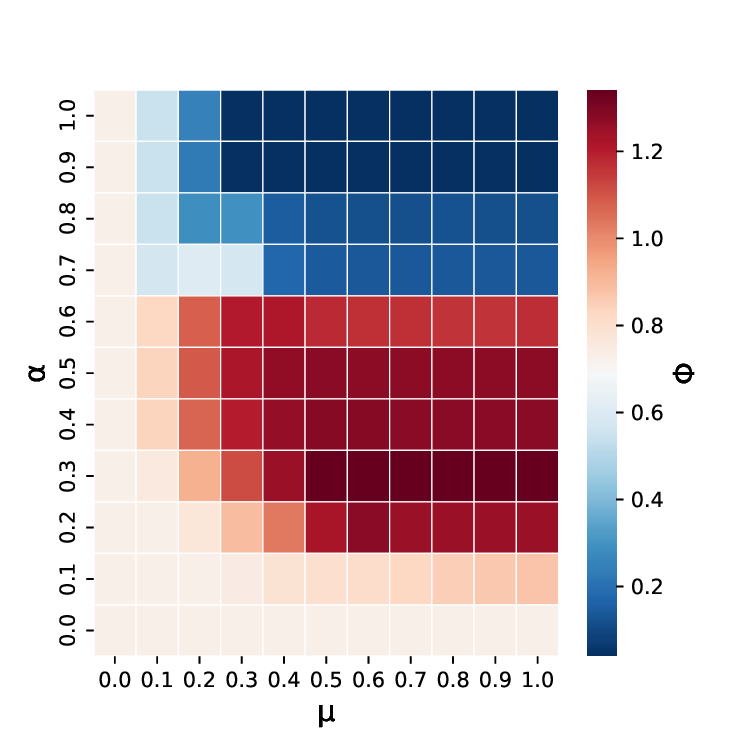}
			\vspace{-1.5\baselineskip}
		\caption{FR, $\beta=1$} 
		\label{fig:FR_PAYOFF}
	\end{subfigure}
	\vspace{-1.5\baselineskip}
	\begin{subfigure}{.9\columnwidth}%
		\includegraphics[width=\columnwidth]
			{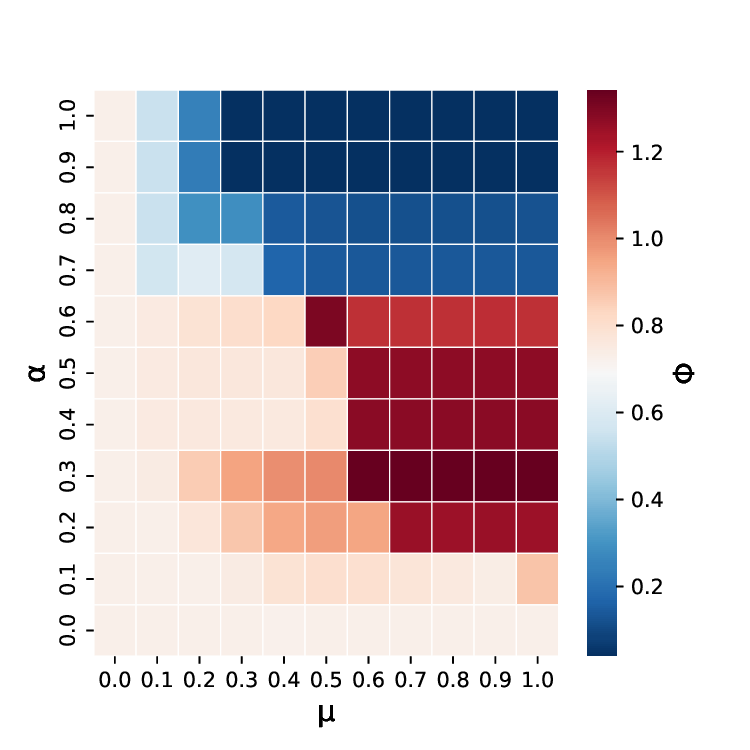}
			\vspace{-1.5\baselineskip}
		\caption{FD, $\beta=1$} 
		\label{fig:FD_PAYOFF}
	\end{subfigure}
	\vspace{-1.5\baselineskip}
	\begin{subfigure}{.9\columnwidth}%
		\includegraphics[width=\columnwidth]%
			{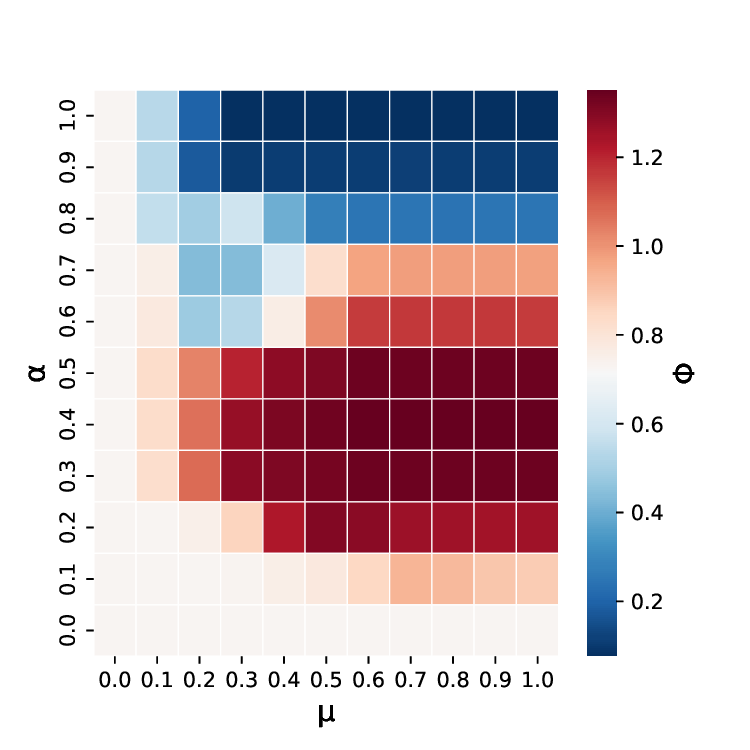}
			\vspace{-1.5\baselineskip}
		\caption{FC, $\beta=2$} 
		\label{fig:payoffs_tol2}
	\end{subfigure}			
	\vspace{-1.5\baselineskip}
	\begin{subfigure}{.9\columnwidth}%
		\includegraphics[width=\columnwidth]%
			{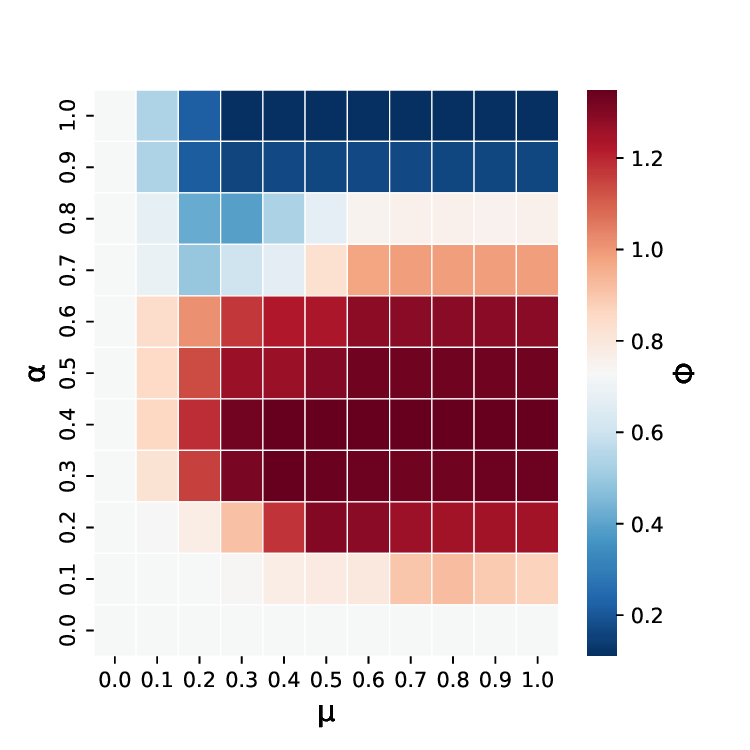}
			\vspace{-1.5\baselineskip}
		\caption{FC, $\beta=3$} 
		\label{fig:payoffs_tol3}
	\end{subfigure}				
	
	\end{multicols}
	\vspace{-1.5\baselineskip}
	\caption{
		Payoff ratios $\phi$ for different 
		memory ratio $\mu$ and 
		optimism threshold $\alpha$ pairs
		for three forgetting strategies (a), (b), (c) and 
		tolerance values (a), (d), (e).
	}
	\label{fig:payoffs}
\end{figure*}

First, we fixed $\beta = 1$ to analyze the effect of optimism threshold $\alpha$ purely. 
The plots in  \reffig{fig:FC_PAYOFF}, \reffig{fig:FR_PAYOFF}, and \reffig{fig:FD_PAYOFF} show the payoff ratios $\phi$ as a function of $\alpha$ and $\mu$ (i.e., the memory ratio $M/N$) with different forgetting strategies. 
We find that cooperators perform better for lower values of $\alpha$ and higher values of $\mu$ in all strategies.

Second, we see that different misjudgments cause different patterns. 
Perceiving a defector as a cooperator is 
safer than perceiving a cooperator as a defector. 
In the first case, 
the agent keeps on playing with the opponent perceived as a cooperator. 
After a couple of defections, 
it realizes that the opponent is a defector and stops playing. 
Hence, the misjudgment is recovered. 
In the second case, there is a wrong punishment situation. 
The blacklisted cooperator cannot obtain a new chance 
to prove that it is not a defector. 
There is no recovery from this misjudgment
unless the agent is forgotten. 
Hence, none of the cooperators can gain payoffs from this matchup.

\begin{figure*}[!tbp]
\begin{multicols}{2}
	\center
	\vspace{-1.5\baselineskip}
	\begin{subfigure}{.9\columnwidth}%
		\includegraphics[width=\columnwidth]
			{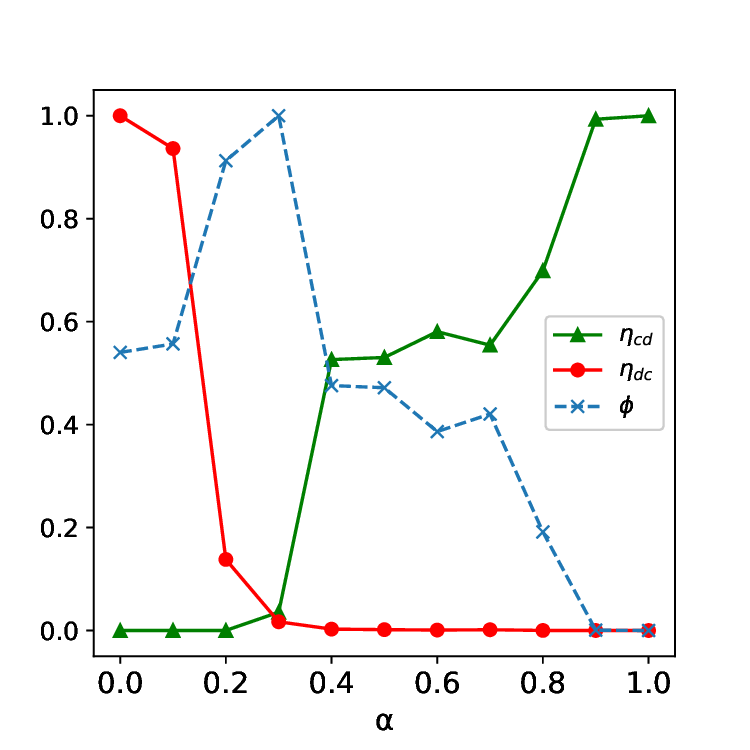}
		\vspace{-1.5\baselineskip}
		\caption{$\mu=0.3$, FC} 
		\label{fig:FC_errors_03}
	\end{subfigure}
	\vspace{-1.5\baselineskip}
	\begin{subfigure}{.9\columnwidth}%
	\vspace{-1.5\baselineskip}
		\includegraphics[width=\columnwidth]
			{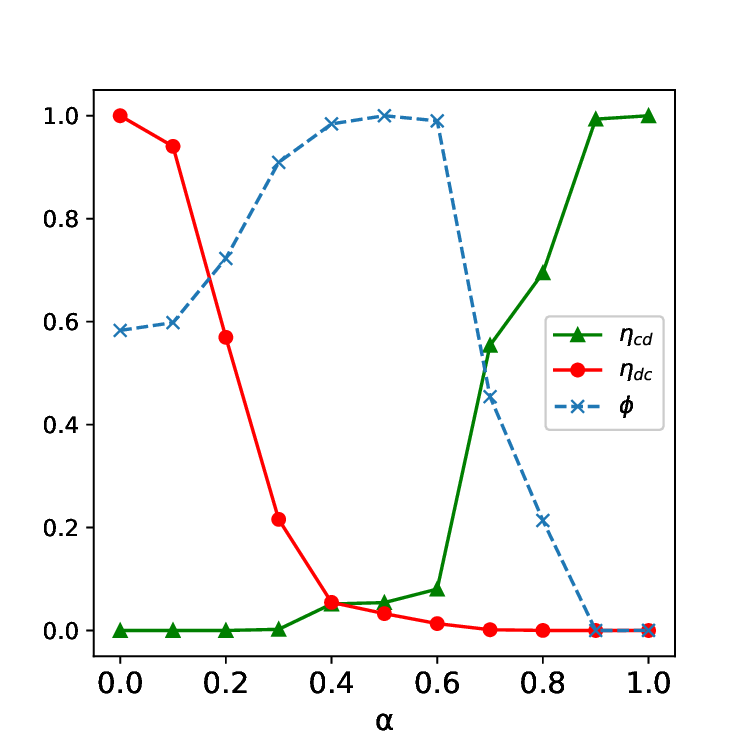}
			\vspace{-1.5\baselineskip}
		\caption{$\mu=0.3$, FR} 
		\label{fig:FR_errors_03}
	\end{subfigure}	
	\vspace{-1.5\baselineskip}
	\begin{subfigure}{.9\columnwidth}%
		\includegraphics[width=\columnwidth]
			{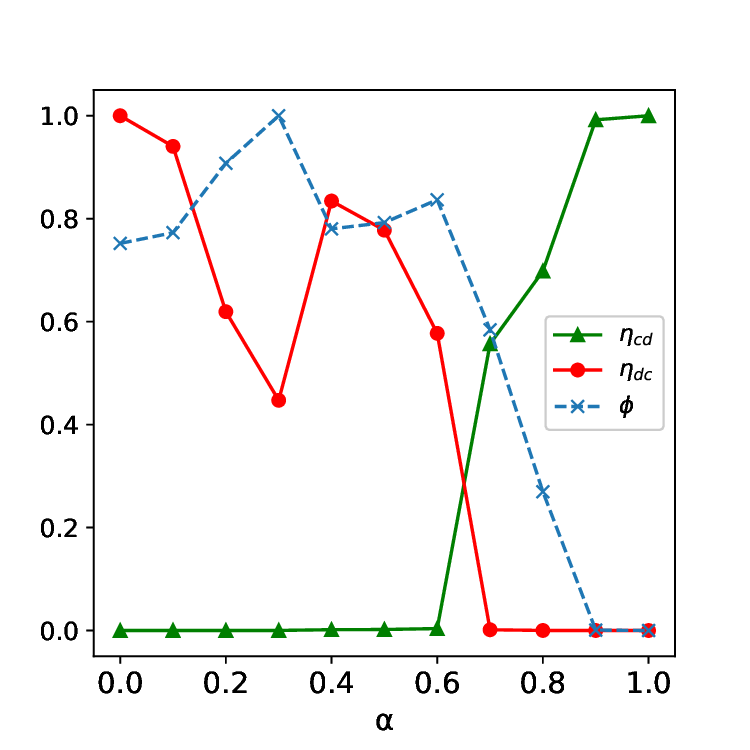}
			\vspace{-1.5\baselineskip}
		\caption{$\mu=0.3$, FD} 
		\label{fig:FD_errors_03}
	\end{subfigure}
	\vspace{-1.5\baselineskip}
	\begin{subfigure}{.9\columnwidth}%
		\includegraphics[width=\columnwidth]
			{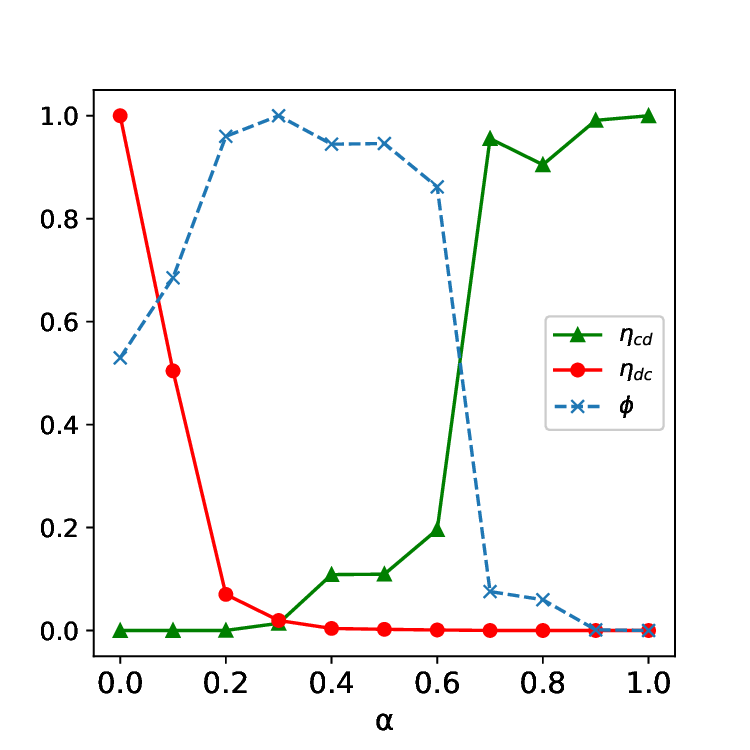}
			\vspace{-1.5\baselineskip}
		\caption{$\mu=0.6$, FC} 
		\label{fig:FC_errors_06}
	\end{subfigure}
	\vspace{-1.5\baselineskip}
	\begin{subfigure}{.9\columnwidth}%
		\includegraphics[width=\columnwidth]
			{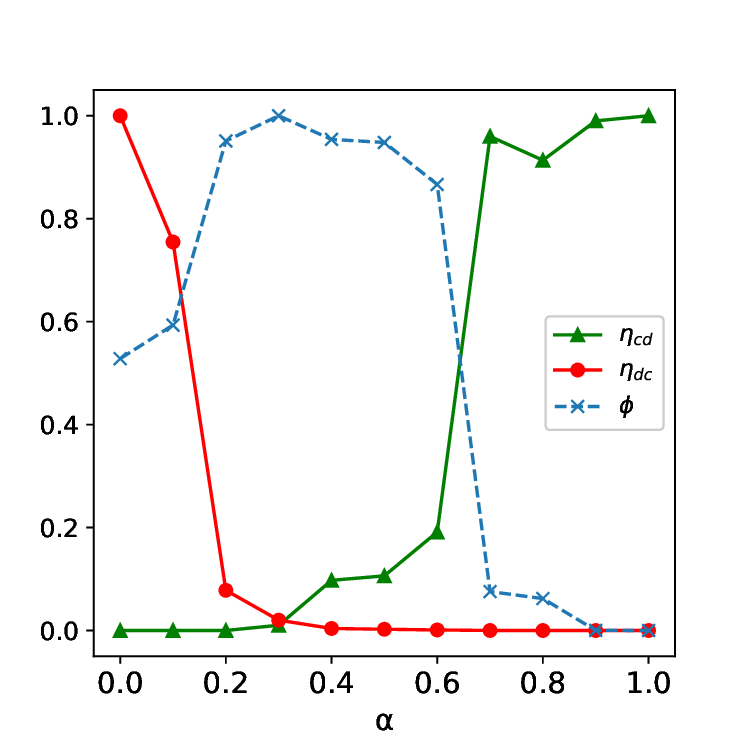}
			\vspace{-1.5\baselineskip}
		\caption{$\mu=0.6$, FR} 
		\label{fig:FR_errors_06}
	\end{subfigure}	
	\vspace{-1.5\baselineskip}
	\begin{subfigure}{.9\columnwidth}%
		\includegraphics[width=\columnwidth]
			{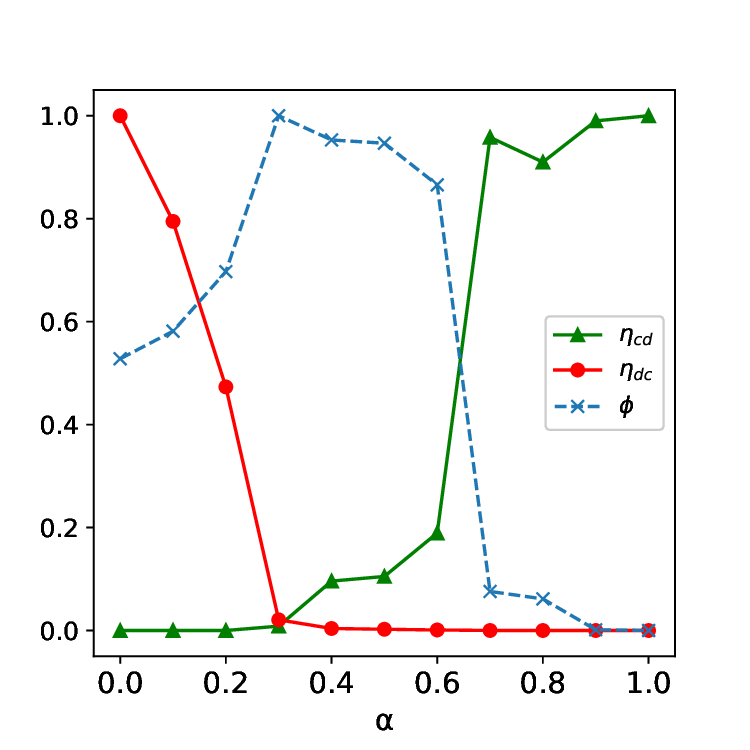}
			\vspace{-1.5\baselineskip}
		\caption{$\mu=0.6$, FD} 
		\label{fig:FD_errors_06}
	\end{subfigure}	
	\end{multicols}
	\vspace{-1.5\baselineskip}
	\caption{
		Normalized number of misjudgments 
		($(\eta - min(\eta))/ (max(\eta) - min(\eta))$)
		for different optimism thresholds $\alpha$ given with the corresponding normalized payoff ratios $\phi$ 
		($(\phi - min(\phi))/ (max(\phi) - min(\phi))$). 
		($\beta = 1$) These results reveal that there is a negative correlation between the number of misjudgments 
		and payoff ratios. 
		Also, high number of $\eta_{cd}$ has a more destructive impact on $\phi$ 
		when compared with high number of $\eta_{dc}$.
	}
	\label{fig:errors1}
\end{figure*}

Wrong punishment brings indirect effects. 
$\alpha$ being low causes the first type of misjudgment, 
whereas $\alpha$ being high causes the second. 
Therefore, low $\alpha$ brings better results than high $\alpha$.  
This finding is comprehensively explained in \refsec{sec:SuccessOfLowOptimismThresholds} with the results in \reffig{fig:errors1}.

Third, FC and FD strategies suffer from misjudged opponents. 
A cooperator labeled as a defector is not forgotten easily in FC strategy. 
In the FD case, the opposite occurs.
A defector labeled as a cooperator is not forgotten easily. 
FR offers a uniform process that eliminates biases 
which leads to more accurate system knowledge. 

If $\beta > 1$, 
as in \reffig{fig:payoffs_tol2} and \reffig{fig:payoffs_tol3}, 
agents have undefined perceptions. 
If most of the agents in memory are perceived as undefined, 
they are forgotten randomly. 
FC and FD turn into FR implicitly in this case. 
This is beneficial because less known opponents 
should not be subject to biased decisions. 
This is shown in \refsec{sec:lowmu} with the results from \reftbl{tbl:anormaly}. 

Although FR has advantages in some cases, 
FC is more successful when it is supported by a low optimism threshold 
or tolerance 
and forgetting biases are prevented. 

Finally, we realized that tolerance has more advantages when $\epsilon \neq 0$ 
since it usually resolves misunderstandings. 
When $\epsilon=0$, there is no need for tolerance since agents can be sure that 
their decision after the first game is definitely correct. 
This finding is investigated in \refsec{sec:highmu}.
Being optimistic was considered to bring vulnerability according to \cite{cinar2020getting}, 
yet there are some advantages as we will explain.

\subsection{Success of Low Optimism Thresholds}
\label{sec:SuccessOfLowOptimismThresholds}

First, we analyzed the results for all optimism thresholds without introducing tolerance,
i.e., $\beta = 1$. 
Note that IPDwRec~\cite{cinar2020getting} worked on $\alpha = 0.5$ case only.
We expand the debate here.
\reffig{fig:FC_PAYOFF}, \reffig{fig:FR_PAYOFF}, and \reffig{fig:FD_PAYOFF} present an overview of the effects of different 
optimism thresholds $\alpha$ combined with different memory ratios. 

The first outcome is the sudden drop in payoff ratios when $\alpha \geq 0.7$. 
The reason is the following: 
Suppose agent $i$ is matched with agent $j$ for the first time, 
and no recommendation is received. 
Then, $i$ has to play. 
Even if $j$ cooperates, 
$(c_{ij}, d_{ij}) = (1, 0)$ leads to $t_{ij} = 2/3 \approx 0.66 < \alpha$. 
That is,
independent of cooperation or defection of $j$, 
$i$ will keep rejecting $j$ until it forgets $j$. 
Therefore, $\alpha \geq 0.7$ represents a meaningless pessimism. 

Second, 
$\alpha \leq 0.1$ means a meaningless optimism
since cooperators fail to be selective. 
For instance,
$i$ will still accept to play with $j$ despite 8 defections, 
since $(c_{ij}, d_{ij}) = (0, 8)$ leads to $t_{ij} = 1/10 = 0.1 \geq \alpha$.

Third, 
cooperators are usually successful when $0.2 \leq \alpha \leq 0.6$. 
The interesting result is that the most successful results are obtained when $\alpha = 0.3$. 
This is going to be investigated.

\subsubsection{$\mu \geq 0.6$}
\label{sec:muOSix}

Since $50\%$ of the agents are defectors,
theoretically, a cooperator with $\mu \geq 0.5$ has enough memory to keep all defectors, 
but this is not the case in practice. 
Cooperators also keep other cooperators in their memories. 
However, we observed that when $\mu \geq 0.6$, agent memories have empty spots at the end of the simulation. 
This is consistent with the results in Ref~\cite{cinar2020getting}. 
Recall that forgetting occurs if memory is full. 
Then forgetting rarely occurs when $\mu \geq 0.6$. 
Therefore forgetting strategies have only minor effects when $\mu \geq 0.6$. 
As expected, the right parts ($\mu \to 1$) of the plots in \reffig{fig:FC_PAYOFF}, \reffig{fig:FR_PAYOFF}, \reffig{fig:FD_PAYOFF} are similar to each other. 

Optimism threshold $\alpha = 0.3$ brings the best results in each of the forgetting strategies in \reffig{fig:payoffs}. 
This is due to the recovery of misjudgments. 
Suppose agent $j$ defected once. 
Then we have $0.3 < t_{ij} = 1/3 < 0.5$. 
Consider $\alpha = 0.3$ and $\alpha = 0.5$ cases. 
(i)~If $\alpha = 0.3$, 
$j$ will get another chance when they are matched again. 
If $j$ cooperates,
$i$ will keep playing with $j$.
If $j$ defects, 
$t_{ij}$ will drop below $\alpha = 0.3$ and 
$i$ will stop playing with $j$. 
Hence, $\eta_{dc}$ will not increase more. 
(ii)~If $\alpha = 0.5$, 
$i$ will not give another chance to $j$. 
If $j$ is a cooperator who defected in the first game, 
its rejection will boost $\eta_{cd}$ and this misjudgment cannot be recovered 
unless $j$ is forgotten. 
Hence, recovery from misjudgment is possible when defectors are perceived as cooperators. 

By looking at 
\reffig{fig:FC_errors_06}, 
\reffig{fig:FR_errors_06}, and 
\reffig{fig:FD_errors_06}, 
where $\mu = 0.6$,
one can see that minimum number of misjudgments appear to be at $\alpha = 0.3$.

\begin{figure}[!bp]
	\center
		\includegraphics[width=\columnwidth]
			{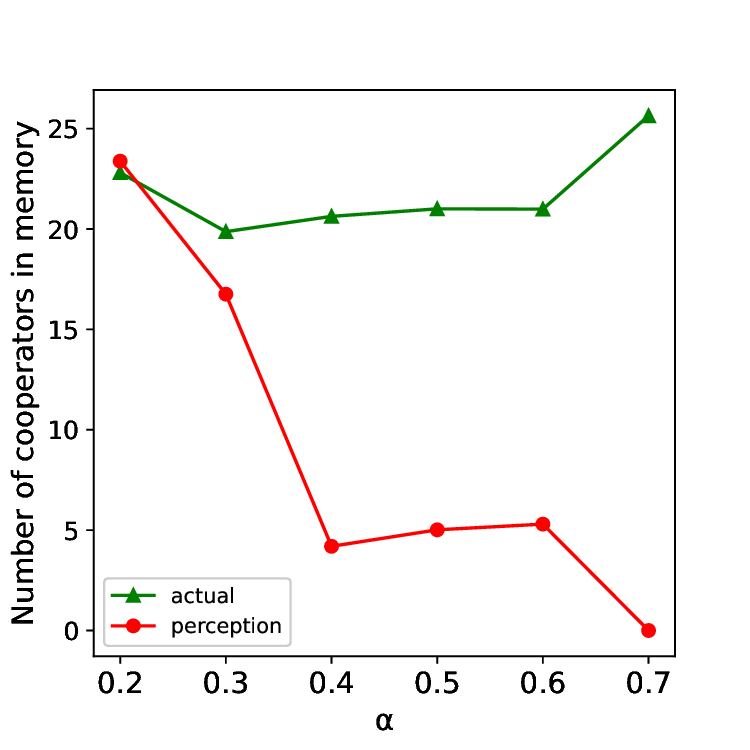}

	\caption{
		Number of cooperators in memory with $\mu=0.3$ for different optimism thresholds $\alpha$,
		in forgetting strategy FC.
		($\beta = 1$)
	} 
	\label{fig:coop_in_memo}
\end{figure}

\subsubsection{$\mu < 0.6$}

For $\mu<0.6$, the effect of optimism thresholds depends on the forgetting strategy.

\textbf{Forget Cooperators.}
As \reffig{fig:FC_PAYOFF} shows, $\alpha = 0.3$ brings the best results for $\mu<0.6$, too. 
To understand this, we collected data from memories of all the cooperators 
at the end of the simulation. 
\reffig{fig:coop_in_memo} shows 
the downside of FC. 
Perceived cooperators are forgotten  
but cooperators perceived as defectors are 
kept in memory. 
The number of perceived cooperators is closer to the number of actual cooperators in low thresholds, 
as seen in \reffig{fig:coop_in_memo}. 
Hence, it is essential to decrease $\alpha$ to decrease $\eta_{cd}$ without increasing $\eta_{dc}$. 
As shown in \reffig{fig:FC_errors_03}, misjudgments are optimized when $\alpha=0.3$.

\textbf{Forget Random.}
Unlike FC strategy, 
FR allows agents to forget independent of their types.
That is, agents in memory have equal chances of being forgotten. 
Hence, cooperators who are perceived as defectors do not remain in memory for a long time, 
which keeps $\eta_{cd}$ under control. 
As observed in \reffig{fig:FR_errors_03}, the optimal tradeoff converges to $\alpha = 0.5$.

\textbf{Forget Defectors.}
FD strategy is dangerous when agents do not keep sufficient defectors in their memories and become vulnerable to them.  
Thus, when $\mu <0.5$, cooperators can not defeat defectors as seen in \reffig{fig:FD_PAYOFF}.

\subsection{Success of Tolerance}

As we explained in \refsec{sec:tol}, the IPDwRec model has no term called tolerance. 
Agents play with their opponent once and decide on the characteristic of the opponent. 
This is the situation when $\beta=1$. 
In this section, we will evaluate the effects of tolerances $\beta=2$ and $\beta=3$. 
For simplicity, FC is going to be used. 
Nevertheless, the effect of tolerance is similar to all forgetting strategies.

\subsubsection{Extreme Thresholds}

In \reffig{fig:FC_PAYOFF}, \reffig{fig:payoffs_tol2}, and \reffig{fig:payoffs_tol3}
payoff ratios $\phi$ are given as a function of memory ratio $\mu$ and optimism threshold $\alpha$ 
for tolerance values $\beta = 1, 2, 3$. 
As previously discussed in \refsec{sec:SuccessOfLowOptimismThresholds}, 
$\alpha > 2/3$ is meaningless  for $\beta = 1$ case 
since even if opponent $j$ cooperates in the first game, $t_{ij}$ becomes $2/3$.
Similar discussion leads that  $\alpha > 3/4$ and  $\alpha > 4/5$ are meaningless 
for  $\beta = 2$ and  $\beta = 3$,
respectively. 
Hence, increasing $\beta$ also shrinks the meaningless region of $\alpha$.

\subsubsection{Low $\mu$ values}
\label{sec:lowmu}

Generally, when agents have low memory ratios, 
cooperators suffer failures due to forgetting biases.
This causes the system to keep inaccurate data. 
Tolerance, by forcing agents to give more chances to their opponents, 
prevents misjudgments generally. 

There are two main reasons for this: 
First, agents tolerate the behaviors in the first games 
and, assuming that $\epsilon$ is small, 
the probability for an agent to make a move, 
which is opposite to its characteristics,
two games in a row is low. 
Second, when agents are matched $k$ times, if $k<\beta$, they perceive each other as undefined, neither as cooperators nor as defectors. 
Then if an agent needs to forget an agent, 
depending on the forgetting strategy, 
it first looks at the list of cooperators or defectors.
If the preferred list is empty, the agent picks an agent from the undefined list to forget. 
Since the undefined list is a random collection of agents, both FC and FD strategies act like FR. 
To sum up, tolerance makes agents uncertain about less known agents, which leads to random forgetting. 
FR strategy eliminates bias.

One anomaly that cooperators encounter can be shown in \reffig{fig:payoffs05FC}. 
When $\beta=1$, and FC strategy is applied at $\mu=0.2$, 
$\phi$ drops significantly to around $0.55$.
Similar drop of $\phi$ around  $\mu=0.2$ is observed in Ref~\cite{cinar2020getting}. 
\reftbl{tbl:anormaly} can be used to study this problem.

\begin{table*}[htp]
\caption{
	Comparison of memory statistics for $\beta=1$ and $\beta=3$. 
	($\alpha =0.5$, $\mu=0.2$, FC)
	\label{tbl:anormaly}
}
\begin{center}
	\begin{tabular}{@{}ccccccccc@{}} 
	\toprule
	$\beta$ & $\phi$ & Actual & Actual  & Perceived  & Perceived & Perceived & $\eta_{cd}$ \\ 
			 &  &cooperators &defectors &cooperators &undefined &defectors &  \\ 
	\hline
	
	1         & 0.5596       & 6.946                              & 13.054                           & 0.739                                
	 & 0                         & 19.260                              & ~~35016.5      \\ 
	3         & 1.1325       & 9.311                              & 10.688                           & 0.05                                 
	 & 8.312                     & 11.637                              & ~~6170.5       \\  
	\bottomrule
	\end{tabular}
\end{center}
\label{default}
\end{table*}

\reftbl{tbl:anormaly} presents statistics from simulations 
where $\alpha = 0.5$ and $\mu=0.2$ for FC.
The values are calculated 
at the end of the simulations by getting data from the memories of cooperators. 
Note that the values in the table are the average values.
Two rows compare the cases of $\beta=1$ and $\beta=3$. 

For $\mu = 0.2$, an agent can keep 20 agents in its memory.
Columns labeled ``actual'' are the numbers of cooperators and defectors kept in the memory, 
which add to 20.
Columns labeled ``perceived'' are the numbers of cooperators, defectors and 
that of the ones considered undefined kept in the memory, 
which also adds to 20.
As expected, for $\beta = 1$, the number of perceived undefined is 0.

The difference in $\eta_{cd}$ statistics implies that 
cooperators refuse to play with each other frequently if $\beta=1$. 
Wrong perception is the reason for this. 
Although there are approximately 7 actual cooperators in the memory,
only 0.739 of them are perceived as cooperators. 
The improvement is clear in the latter row where $\beta=3$. 
Although almost no one is perceived as a cooperator, 
approximately 8 agents are kept as undefined in memory, 
which means those 8 agents will not be rejected when matched. 
Being tolerant moved cooperators that were perceived as defectors to the undefined list, 
which can be seen as giving more chances to the agents 
who are not surely defectors.

\subsubsection{High $\mu$ values}
\label{sec:highmu}

\begin{figure}[!tbp]
	\center
	\begin{subfigure}{\columnwidth}%
	\includegraphics[width=\columnwidth]%
		{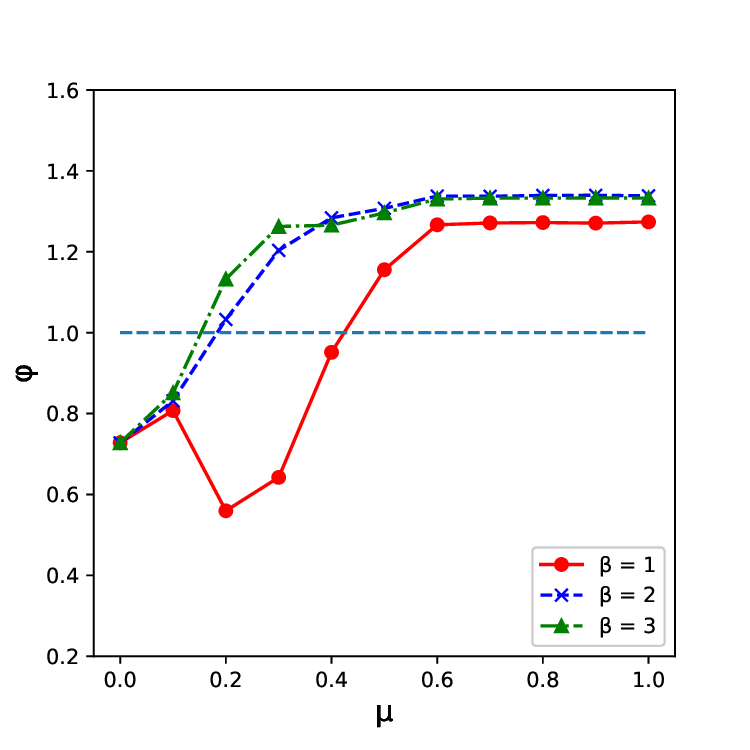}
		\caption{$\epsilon=0.1$} 
		\label{fig:payoffs05FC}
	\end{subfigure}	
	
	\begin{subfigure}{\columnwidth}%
	\includegraphics[width=\columnwidth]%
		{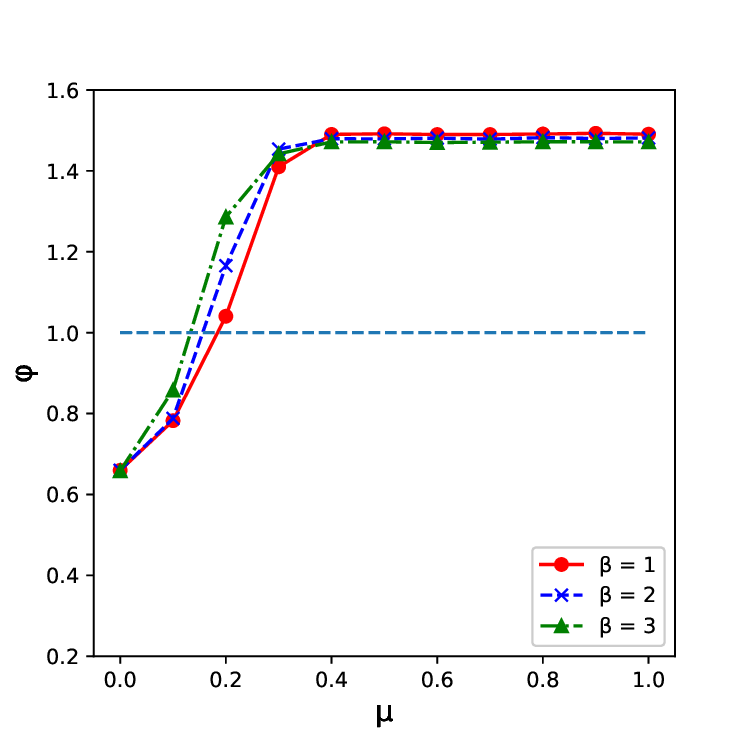}
		\caption{$\epsilon=0$} 
		\label{fig:payoffs05FCnoerror}
	\end{subfigure}		
		
	
	\caption{
		Payoff ratios $\phi$ for different 
		memory ratios $\mu$
		for three tolerance values $\beta$,
		where optimism threshold $\alpha = 0.5$ and FC applied.
	} 
	\label{fig:payoffs05}
\end{figure}

\reffig{fig:payoffs05FC} shows $\phi$ as a function of $\mu$ when $\beta = 1, 2, 3$.
The payoffs for $\beta = 2$ and $3$ are higher than that of $\beta = 1$ for high values of $\mu$, too.
Even though there is no bias due to forgetting or lack of memory when $\mu \geq 0.6$, 
being tolerant is better. 
This is due to the following: 
Since $\epsilon = 0.1$,
approximately $10~\%$ of the cooperators defect and $10~\%$ of the defectors cooperate in the first game.
After that, their opponents will always perceive those agents incorrectly when $\beta=1$. 

To test this hypothesis, we can take a look at the results in \reffig{fig:payoffs05FCnoerror} 
where $\epsilon = 0$.
This means that cooperators are pure cooperators who never defect, 
and defectors are those who never cooperate. 
It can be claimed that tolerance has no benefit when $\mu \geq 0.6$ and $\epsilon = 0$. 
Then we conclude that tolerance repairs the issues due to mixed strategy when $\mu \geq 0.6$.

\begin{figure}[!tbp]
	\center
		\includegraphics[width=\columnwidth]
			{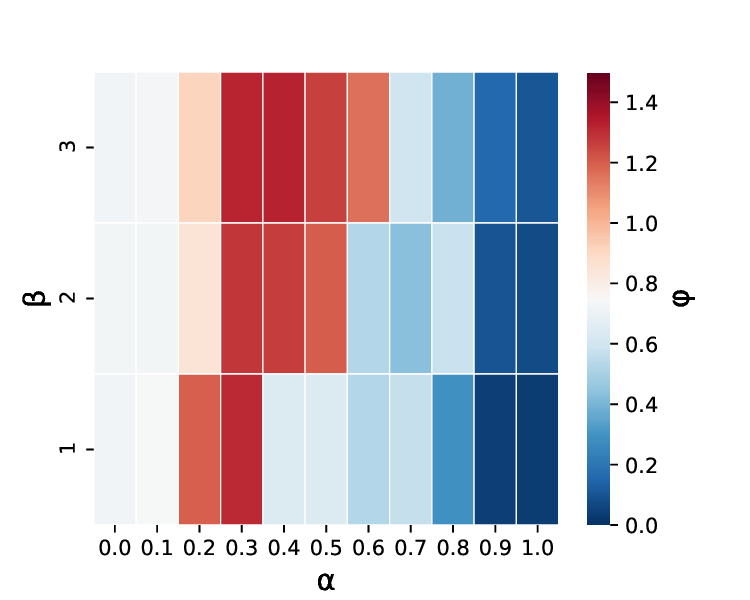}
	
	\caption{
		Payoff Ratios $\phi$ for tolerance $\beta$ and optimism threshold $\alpha$ pairs 
		when FC applied and $\mu = 0.3$.
	} 
	\label{fig:tradeoff}
\end{figure}

\subsection{Tradeoff Between Threshold and Tolerance}

Both optimism threshold and tolerance affect the dispositions and performances of cooperators. 
Having a low optimism threshold can be interpreted as being optimistic as cooperators give second chances to their opponents even if they face a defection in the first game. 
Having high tolerance also makes agents more optimistic because agents keep the hope that the opponent can still be a cooperator until tolerance ends.  

Although being optimistic with these terms increased the performance by eliminating biases and incorrect perceptions, 
applying both of them did not turn up to be successful after a while. 
\reffig{fig:tradeoff} shows that there is a tradeoff between increasing $\beta$ and decreasing $\alpha$. 
The red areas show high success, 
and the positive sloped pattern of red areas indicates that optimism threshold and tolerance should be considered together. 
If $\alpha$ is too low, increasing $\beta$ can destroy the performance of cooperators. 
Hence, to increase $\beta$, $\alpha$ should also be increased. 
Being more optimistic in the sense of tolerance requires being more pessimistic in the sense of threshold.

\section{Future Work}

This work aimed to provide new disposition parameters $\alpha$ and $\beta$ and analyze the effects. 
We kept those parameters the same within the simulation and compared simulations. 
To reach more concrete conclusions, 
an evolutionary approach can be applied with a system in which agents differ in $\alpha$ and $\beta$. 
Optimal $\alpha$ and $\beta$ can be obtained with surviving parameter. 
Moreover, agents can change their own $\alpha$ and $\beta$ in time 
depending on conditions such as memory status or the number of games played. 
This flexibility may lead to a more successful performance. 

We imported the same forgetting strategies as in Ref~\cite{%
	cinar2020getting} 
to our model. 
Different forgetting strategies can also be considered. 

Additionally, the effects of system parameters as $\tau$, $N$, and $\epsilon$ can be investigated. 
If we increase $\tau$, the game becomes longer, and higher $\beta$ may be required. 
Also, we observed the difference between $\epsilon = 0$ and $\epsilon = 0.1$. 
If $\epsilon$ becomes higher, optimal $\alpha$ and $\beta$ values may change. 

In our model, the agents do not keep track of recommendation behaviors of each other.
If they do,
as  pointed out by one of our reviewers, 
cooperators can suspect that an agent, 
who does not response to calls, 
maybe a defector.
With a similar deduction,
one suspect that an agent, 
who does not asks for recommendation, 
maybe a defector.

In both IPDwRec and our model, 
agents provide genuine information that they have.
One may extend the model by introducing manipulated recommendations.
For instance,
answering with a low value of the perceived cooperation probability,
although it actually calculates a high value.

\section{Conclusion}

We introduced new parameters to Iterated Prisoner's Dilemma game with limited memory
to inject more precise dispositions into the system. 
In addition, defectors always play in this model and do not provide recommendations.
This makes the game more difficult for cooperators compared to the previous studies.
Yet, cooperators can be successful in getting higher average payoff ratios than defectors 
if they are more optimistic.
Namely, if they forgive the first few defections by means of tolerance, optimism threshold, or both. 
We related this result to misjudgments.
That is, perceiving a cooperator as a defector is more dangerous than perceiving a defector as a cooperator. 
Moreover, FR strategy is better against less known agents since FC and FD strategies cause bias, 
which may be unfair if the agents are not well known. 
Since even a small probability of impurity in characteristic ($\epsilon = 0.1$) 
results in incurable misjudgments, 
acting tolerant and unbiased is necessary for cooperators to obtain a more accurate perception.

\section*{Acknowledgements}

We would like to thank our anonymous reviewers for their elaborative feedback. 
We also would like to thank Meliksah Turker, Bora Dogan, and Mehmet Can Turkes for their constructive comments.  
This work is partially supported by the Turkish Directorate of Strategy and Budget 
under the TAM Project number 2007K12-873.

\textbf{Code.}
The code to run the simulation and test other configurations is available at \url{https://github.com/orhungorkem/IPDwithTolerance}

\bibliographystyle{ieeetr}
\bibliography{IPDwR-Rel-v03}

\end{document}